\documentclass{moriond}

\def\beq{\begin{equation}}
\def\eeq{\end{equation}}
\def\bea{\begin{eqnarray}}
\def\eea{\end{eqnarray}}
\def\beqn{\begin{eqnarray}}
\def\eeqn{\end{eqnarray}}

\def\alphas{\alpha_{\rm S}}
\def\as{a_{\rm S}}

\newcommand{\Photo}{}

\begin{document}
\vspace*{3.7cm}
\title{Including higher-order mixed QCD-QED effects \\ in hadronic calculations}

\author{Germ\'an F. R. Sborlini}

\address{Tif lab, Dipartimento di Fisica, Universit\`a di Milano and INFN, Sezione di Milano, \\
I-20133 Milan, Italy.}

{\tiny TIF-UNIMI-2018-5}

\maketitle\abstracts{
We review some recent results about the computation of mixed QCD-QED corrections beyond the leading order in perturbation theory. We start by considering the effects induced in the Altarelli-Parisi equations and the partonic distributions. In particular, we describe the computation of one-loop mixed QCD-QED and two-loop QED terms in the splitting kernels, which are relevant to account for the presence of photon distributions. In the last part, we briefly talk about the implementation of these corrections in the context of the $q_T$-subtraction/resummation formalism.}

\section{Introduction and motivation}
The study of hadronic collisions in the high-energy and high-luminosity regime is pushing the theoretical predictions to the precision frontier. The experimental analysis require accurate simulations which should include most of the effects predicted by the theoretical framework. With more data becoming available, the experimental uncertainties rapidly reduce and this increases the sensitivity to tiny deviations from the dominant QCD background. In particular, the QED and electroweak contributions are turning into crucial components of the full predictions since they could provide percent-level modifications, now detectable by the experiments.

The purpose of this work is to briefly summarize some recent developments concerning the precision program in QCD-QED computations. In Sec. \ref{sec:splittings}, we describe the extension of the DGLAP equations to include mixed ${\cal O}(\alpha \alpha_S)$ corrections. In Sec. \ref{ssec:abelianization}, we comment on a technique that allows to recover these QED and mixed QCD-QED contributions from the well-known QCD calculations. After that, in Sec. \ref{sec:difotones}, we apply the Abelianization algorithm to a particular collider process, namely diphoton production, and we show that the Abelianized $q_T$-subtraction framework successfully deals with the singularities due to the presence of soft/collinear gluons and photons. Finally, we present the conclusions in Sec. \ref{sec:conclusions}.


\section{Extended DGLAP equations}
\label{sec:splittings}
The DGLAP equations control the perturbative evolution of the parton distribution functions, through the well-known splitting kernels. These objects can be computed within perturbation theory, by studying the collinear limits of scattering amplitudes. Higher-order corrections to the splitting functions for QCD partons have been computed, both in the multi-loop and in the multiple-collinear case. In particular, we have explored the double and triple collinear limits at one-loop level for processes including photons\cite{Splittings}.

The formalism can be extended to include leptons and photons, as well as QCD partons. This leads to a more complicated system of integro-differential coupled equations which can be partially simplified by changing the PDF basis. Explicitly, by using the canonical basis, i.e. ${\cal B}=\{u,\bar u, d,\bar d, \ldots, e^-, e^+,\ldots, \gamma, g\}$, the system is written as
\beq
\frac{d f_i}{dt} = \sum_{j \in {\cal B}} P_{f_i f_j} \otimes f_j \, ,
\eeq
with $t=\log(Q^2/\mu^2)$ the evolution variable and $\otimes$ denotes a convolution among distributions ($f_i$) and splitting kernels ($P_{f_i f_j}$). The last ones are expanded in a perturbative series according to 
\beq
P_{ij} = \sum_{k,l} \, \as^k a^l \, P_{ij}^{(k,l)} \, ,
\eeq
with $\alpha \equiv 2\pi \, a$ and $\alphas \equiv 2\pi \, \as$. In general, the evolution kernels are not diagonal, thus conducing to a highly non-trivial mixing of the different distributions. However, by using an optimized basis\cite{Roth:2004ti}, we get 
\beq
\frac{d \{\Delta_{2}^l,\Delta_{3}^l\}}{dt} =  P_{l}^+ \otimes \{\Delta_{2}^l,\Delta_{3}^l \}  \, , \quad \frac{d \{ \Delta_{U} \}}{dt} =  P_{u}^+ \otimes \{ \Delta_{U}  \} \, , \quad \frac{d \{ \Delta_{D}  \}}{dt} =  P_{d}^+ \otimes \{ \Delta_{D}  \} \, ,
\label{eq:diagonales}
\eeq
with $\Delta_i^l$, $\Delta_U$ and $\Delta_D$ some specific linear combinations of leptons, up and down quarks PDFs, respectively\cite{deFlorian:2015ujt,deFlorian:2016gvk}. Moreover, up to ${\cal O}(\alpha \, \alphas^n)$, the kernels relating different quark and/or lepton flavors vanish, which allows to write the equations for the valence distributions as
\beq
\frac{dq_{v_i}}{dt} =   P_{q_i}^-     \otimes q_{v_i}    \, , \quad \quad \frac{dl_{v_i}}{dt} =   P_{l}^-     \otimes l_{v_i}  \, , \quad \quad u_{v} = u - \bar u, \ldots, e_{v}=e^- - e^+ , \ldots \
\label{eq:diagonalVALENCIA}
\eeq
besides highly simplifying the singlet equations\cite{deFlorian:2015ujt,deFlorian:2016gvk}.

The full set of evolution equations must fulfill physical constraints in the end-point region (i.e. $x=1$), which are related to the conservation of the proton total momentum and its fermionic composition (i.e. the fermion number conservation). These conditions are used to derive the terms proportional to $\delta(1-x)$ in the splitting functions, besides imposing strong restrictions on the functional dependence of the different kernels for $0 \leq x < 1$.


\subsection{Abelianization of splitting kernels}
\label{ssec:abelianization}
The key idea of our work is related with the possibility of extracting the mixed QCD-QED corrections from previous computations performed within pure QCD. The algorithmic approach is based on the Abelianization of the different contributions to the QCD computation, which accounts for the replacement of \emph{gluons by photons}\cite{deFlorian:2015ujt}. If the computation is available on a diagram-by-diagram decomposition, then the Abelianization is trivial: we can strip the color factor and multiply the kinematic part by the recomputed factor, including the charge dependence. But, the interesting fact is the possibility of recovering the QCD-QED corrections from \emph{analytic results} provided that we keep track of the color factors (i.e. $C_A$, $C_F$) and contributions proportional to $n_F$. Then, we can derive explicit rules to transform the results. For instance, starting from the ${\cal O}(\alphas^2)$ splitting functions, the ${\cal O}(\alpha \, \alphas)$ terms are obtained by:
\begin{itemize}
\item removing the color average over the initial states, writing the color factors as functions of $N_C$ and expanding around $N_C=0$;
\item keeping the leading terms and recomputing the color structure (inserting the electric charges where it corresponds). 
\end{itemize}
There are some subtleties related with the presence of symmetry factors and the treatment the closed fermion loops: we need to consider the replacement $n_F \to \sum_q e_q^2$. In an analogous way, well-defined transition rules to extract the ${\cal O}(\alpha^2)$ contributions can be defined\cite{deFlorian:2016gvk}.  

\begin{figure}[htb]
\begin{center}
\includegraphics[width=0.80\textwidth]{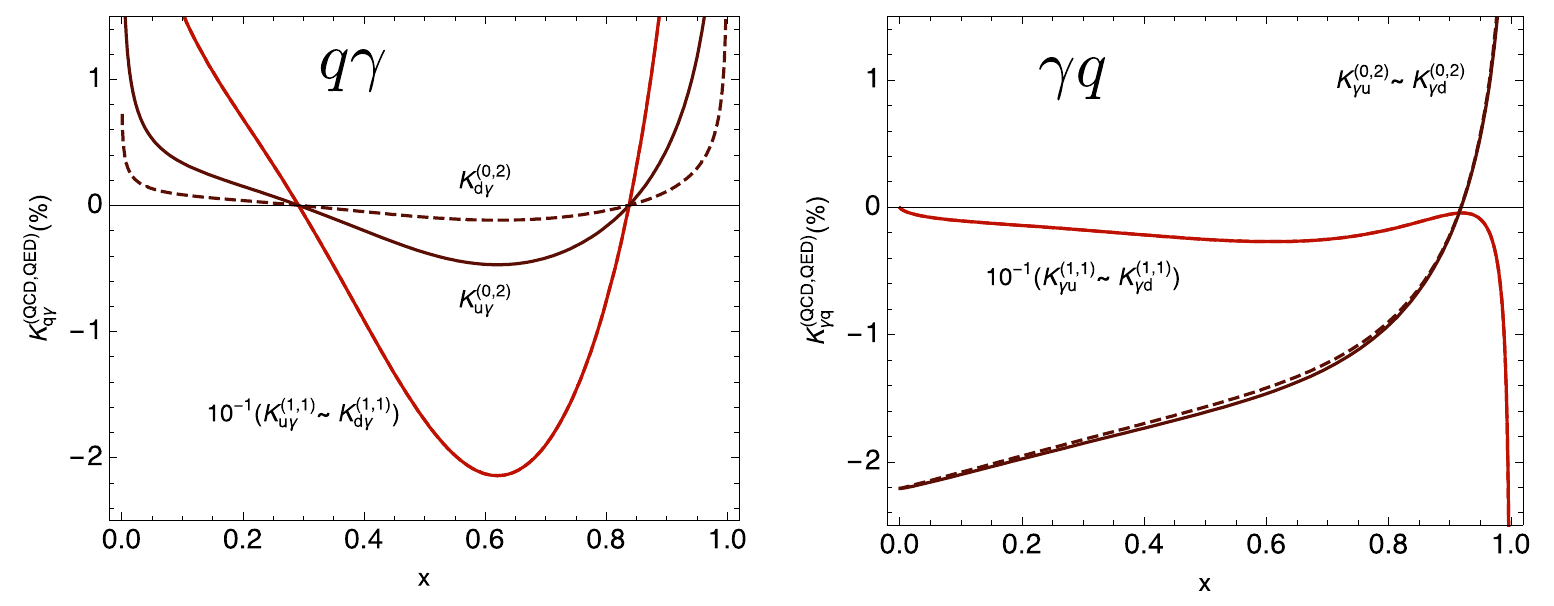}
\caption{Corrections due to the inclusion of QED contributions in the $P_{q\gamma}$ (left) and $P_{\gamma q}$ splitting kernels. We include both ${\cal O}(\alpha^2)$ (brown) and ${\cal O}(\alpha\,  \alphas)$ (red) terms.} 
\label{fig:Kernelsqgamma}
\end{center}
\end{figure} 

To quantify the impact of the QED corrections, we define the $K$-ratios according to
\beq
K^{(i,j)}_{ab} = \as^i \, a^j \, P^{(i,j)}_{ab} (x) / P^{\rm LO}_{ab}(x) \, , \quad \quad P^{\rm LO}_{ab} = \as \,  P^{(1,0)}_{ab} + a \, P^{(0,1)}_{ab} \, ,
\label{eq:K-ratios}
\eeq
where the lowest-order kernel contains both the lowest QED and QCD contributions. It is important to notice that these corrections allow to connect the quark and photon PDFs, thus leading to a non-trivial evolution of the last distribution. In Fig. \ref{fig:Kernelsqgamma}, we plot the $K$-ratios for the $P_{q\gamma}$ and $P_{\gamma q}$ splitting kernels. The ${\cal O}(\alpha^2)$ (brown) contributions are suppressed by a factor 10 in comparison to the dominant ${\cal O}(\alpha \, \alphas)$ ones (red). A non-trivial charge separation effect is present in the middle-$x$ region, in particular for the $P_{q\gamma}$ kernel.


\section{Application to diphoton production}
\label{sec:difotones}
By applying the Abelianization algorithm to the public code \texttt{2gNNLO} \cite{Catani:2011qz}, we modified the structure of the NLO QCD terms to recover the corresponding NLO QED contributions. In this procedure, we obtained the NLO QED extension of the $q_T$-subtraction formalism\cite{Catani:2007vq,Sborlini:2017gpl}, which allows to consistently deal with the fixed-order calculation beyond the LO in QCD-QED. Moreover, we checked the behavior of the different pieces of the calculation and found a complete cancellation of IR singularities in the $q_T \to 0$ limit. 

\begin{figure}[htb]
\begin{center}
\includegraphics[width=0.506\textwidth]{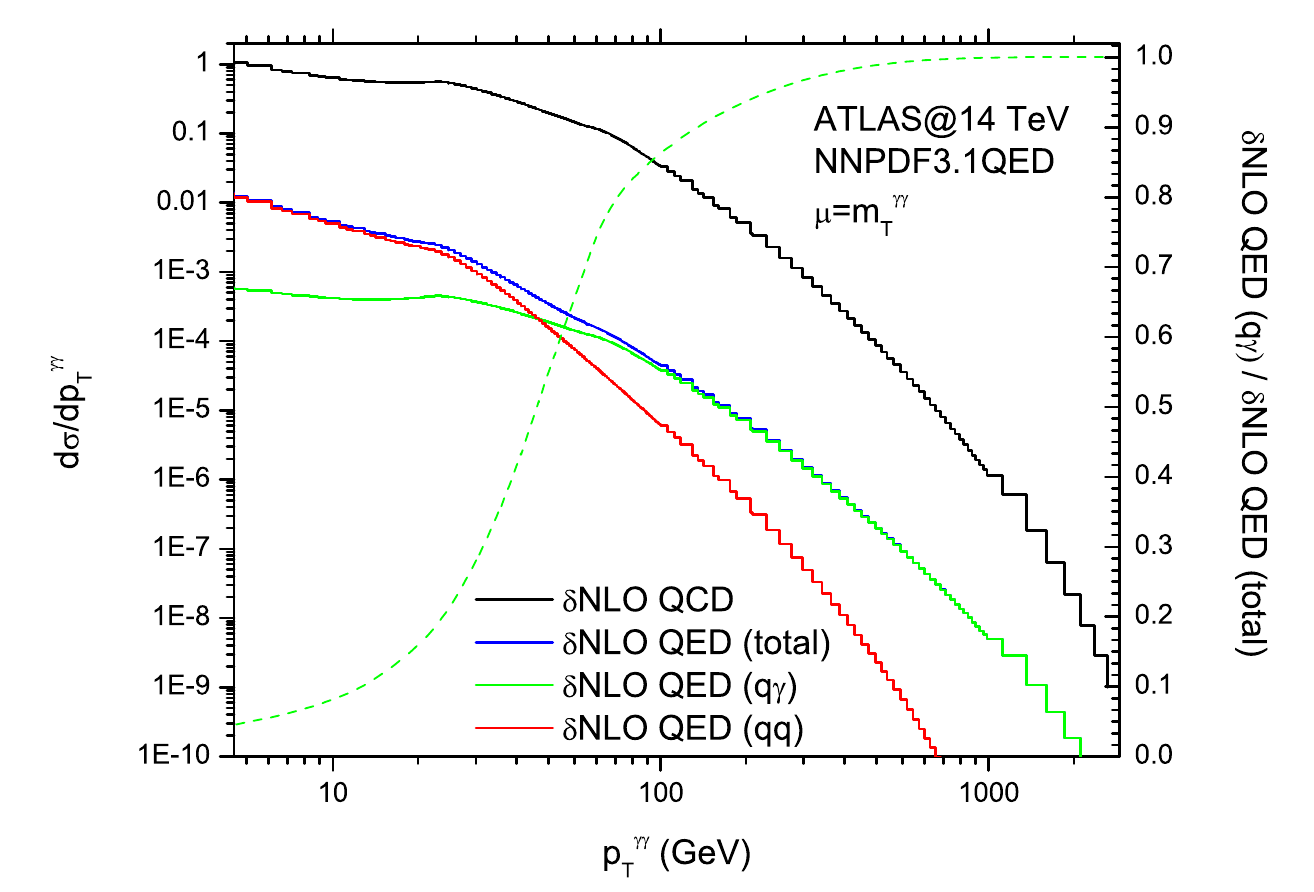} \ \
\includegraphics[width=0.455\textwidth]{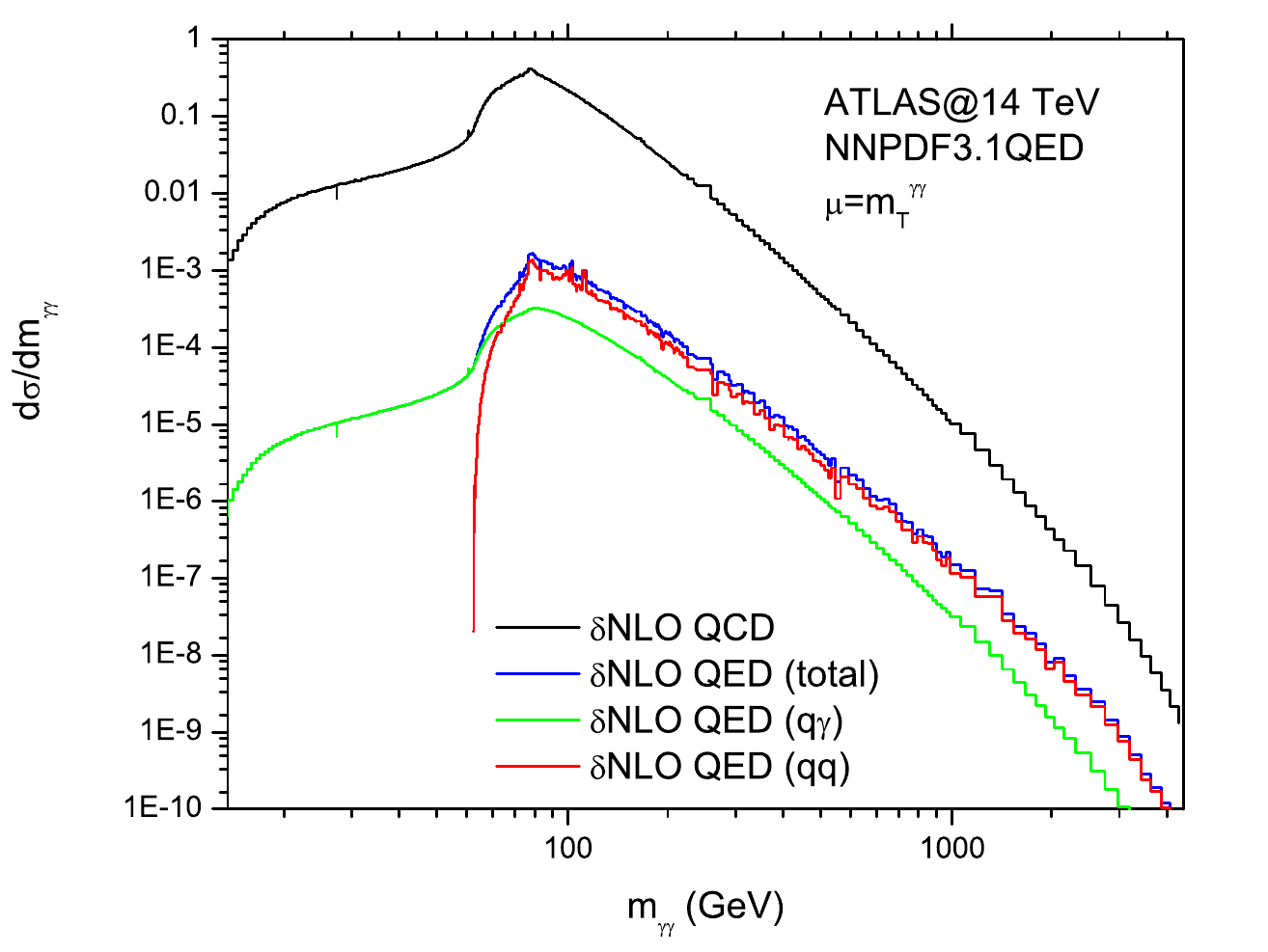} 
\caption{Transverse momentum (left) and invariant mass (right) distributions for the diphoton production at LHC. The black (blue) curve shows the total NLO QCD (QED) prediction, without including the LO. The dashed green line indicates the relative contribution of the $q \gamma$-channel to the total NLO QED correction.}
\label{fig:Difotones}
\end{center}
\end{figure}

In Fig. \ref{fig:Difotones}, we present the NLO QED corrections to the transverse momentum (left) and invariant mass (right) distributions for the diphoton production at LHC. We used the typical ATLAS cuts, with 14 TeV c.m. energy and \texttt{NNPDF3.1QED}~\cite{Ball:2017nwa,Bertone:2017bme} PDFs (which implements the \texttt{LUXqed} approach\cite{LUXQED} to reduce the uncertainties in the determination of the photon PDF). In the invariant-mass spectrum, the momentum ordering of the triple-photon system leads to a dynamical cut, which is a novel feature of the QED corrections. Also, we notice that the $q\gamma$-channel (green line) dominates the NLO QED contribution in the high-transverse momentum region, thus showing the importance of a precise determination of the photon PDF.

\section{Conclusions and outlook}
\label{sec:conclusions}
In this article, we summarized some recent developments in the computation of mixed QCD-QED corrections. We provided an algorithmic technique to recover these contributions from the higher-order QCD computations. By applying the Abelianization procedure, we managed to obtain the ${\cal O}(\alpha \, \alphas)$~\cite{deFlorian:2015ujt} and ${\cal O}(\alpha^2)$~\cite{deFlorian:2016gvk} corrections to the Altarelli-Parisi splitting functions. This knowledge is crucial for the proper treatment of the PDF evolution beyond LO QED.

Besides that, the Abelianization proved to be useful for obtaining corrections to physical cross-sections. We applied it through the $q_T$-subtraction method, and obtained the universal coefficients to compute the NLO QED corrections to the production of neutral colorless particles. As a proof-of-concept, we included ${\cal O}(\alpha^3)$ terms to the diphoton production cross-section~\cite{Sborlini:2017gpl,INPREP}. We discussed the phenomenology for hadron colliders (LHC at $14 \, {\rm TeV}$), and we found non-negligible corrections in the high-energy region. 

Further considerations on the development of a framework to deal with mixed QCD-QED is needed to achieve precise predictions for current experiments\cite{INPREP}. In this direction, we have managed to successfully extend the $q_T$-resummation formalism to compute mixed QCD-QED corrections, using on-shell $Z$ production as benchmark process\cite{INPREP2}.


\section*{Acknowledgments}
This work has been done in collaboration with D. de Florian, G. Rodrigo, L. Cieri and G. Ferrera. The research project was supported by CONICET, ANPCyT, the Spanish Government, EU ERDF funds (grants FPA2014-53631-C2-1-P and SEV-2014-0398) and Fondazione Cariplo under the grant number 2015-0761. This article is based upon work from COST Action CA16201 PARTICLEFACE supported by COST (European Cooperation in Science and Technology).

\end{document}